\documentclass[12pt]{article}
\usepackage{amssymb}

\usepackage{graphicx}

\advance \textheight by \topskip
\newcommand{\be}[1]{\begin{equation}\label{#1}}
\newcommand{\beq}{\begin{equation}}
\newcommand{\ee}{\end{equation}}
\newcommand{\beqn}[1]{\begin{eqnarray}\label{#1}}
\newcommand{\eeqn}{\end{eqnarray}}

\begin{document}

\title{Baryogenesis: The Lepton Leaking Mechanism}
\author{\underline{Lu\'{\i}s Bento}\thanks{
Talk given at the
\emph{XI International School: Particles and Cosmology}, 
Baksan Valley, Kabardino-Balkaria, Russia, 18-24 April, 2001.}
 \\
{\normalsize \emph{Centro de F\'{\i}sica Nuclear da Universidade de Lisboa, }
} \\
{\normalsize \emph{Avenida Prof. Gama Pinto 2, 1649-003 Lisboa, Portugal }} 
\\
\\
Zurab Berezhiani \\
{\normalsize \emph{\ Dipartamento di Fisica, Universit\'a di L'Aquila,
I-67010 Coppito, AQ, Italy }} \\
{\normalsize \emph{\ INFN, Laboratori Nazionali del Gran Sasso, I-67010
Assergi, AQ, Italy }} \\
{\normalsize \emph{\ Andronikashvili Institute of Physics, GE-380077
Tbilisi, Georgia }}}
\date{}
\maketitle

\begin{abstract}
We propose a baryo- and leptogenesis mechanism in which the $B-L$ \
asymmetry is produced in the conversion of ordinary leptons into particles
of some depleted hidden sector. In particular, we consider the lepton number
violating reactions $l\phi \rightarrow l^{\prime }\phi ^{\prime },\,\bar{l}%
^{\prime }\bar{\phi}^{\prime }$ mediated by the heavy Majorana neutrinos $N$
of the seesaw mechanism, where $l$ and $\phi $ are ordinary lepton and Higgs
doublets and $l^{\prime }$, $\phi ^{\prime }$ the ``sterile'' leptons and
Higgs. This mechanism can operate even if the reheat temperature is
smaller than the $N$ Majorana masses, in which case the usual leptogenesis
mechanism through $N$ decays is ineffective. In particular, the reheat
tempearture can be as low as $10^{9}$ GeV or less.
\end{abstract}

\textbf{1. Baryogenesis through Leptogenesis}

\strut

It is well known that in order to produce a non-zero baryon asymmetry from
the initially baryon symmetric Universe three conditions must be fulfilled~ 
\cite{sakharov67}: 1) $B$-violation, 2) $C$ and $CP$ violation and 3)
departure from thermal equilibrium. These conditions can be satisfied in the
decays of heavy gauge or Higgs bosons in the context of grand unification.
In the standard model $B$ and $L$ are also violated by electroweak instanton
(sphaleron) processes~\cite{kuzmin85} however they are in thermal
equilibrium at temperatures from about $10^{12}$ GeV down to the electroweak
phase transition at $100$ GeV. Thus they can potentially erase the
primordial baryon number. In fact they still conserve $B-L$ like all the
other standard model interactions and it has been shown~\cite
{khlebnikov88,harvey90,Dolgov} that under thermal equilibrium conditions the
equations of detailed balance constrain the baryon and lepton asymmetries to
be proportional to each other or better, to $B-L$: 
\begin{equation}
B=C(B-L)\;,\qquad L=(C-1)(B-L)\;.
\end{equation}
$C$ is an order one coefficient that depends on which interactions and set
of particles are in chemical equilibrium and for that reason varies with the
temperature scale. For instance, in the standard model with three fermion
families and one Higgs doublet, $C=28/79$ before and $C=36/111$ during the
electroweak phase transition. Therefore, the present baryon to entropy
ratio, $Y_{B}=n_{B}/s=(0.6-1)\times 10^{-10}$ implies a $B-L$ asymmetry $%
Y_{B-L}\sim (2-3)\times 10^{-10}$.

This tells two things. One is that one actually needs to violate and produce
a non-zero $B-L$ number and not just $B$. This disfavours the simplest
picture based on grand unification models like $SU(5)$ where $B-L$ is
conserved. The other is that it is not essential to have explicit baryon
number violating interactions. A $B-L$ asymmetry may be generated thanks to 
$L$ violating, $B$ conserving interactions while the $B-L$ asymmetry is
transported to the baryon sector through electroweak instantons. The first
realization of such idea, generically called leptogenesis, is the delayed
heavy neutrino decay~\cite{fukujita86,luty92,buchmuller99}.

\strut

\textbf{2. Delayed Heavy Neutrino Decay}

\strut

In this mechanism, directly related to the seesaw scheme for the light
neutrino masses~\cite{seesaw}, $B-L$ is generated in the decays of heavy
Majorana neutrinos, $N$, into leptons $l$ and anti-leptons $\bar{l}$ ($\phi $
is a standard Higgs doublet) 
\begin{equation}
N\rightarrow l\,\phi \,,\,\bar{l}\,\bar{\phi}\;.
\end{equation}
In this context, the three necessary Sakharov conditions are realized in the
following way. 1) $B-L$ and $L$ are violated by the heavy neutrino Majorana
masses. 2) The out-of-equilibrium condition is satisfied thanks to the delayed
decay(s) of the Majorana neutrinos i.e., decay rate(s) smaller than the
Hubble rate, 
\begin{equation}
\Gamma _{N}<H\qquad (T\sim M_{N})\;,
\end{equation}
or, life-time(s) larger than the age of the Universe at the time they become
non-relativistic. 3) The origin of $CP$-violation ($C$ is trivially violated
due to the chiral nature of the fermion weak eigenstates) are the $lN\phi $
complex Yukawa couplings resulting in asymmetric decay rates 
\begin{equation}
\Gamma (N\rightarrow l\phi )\neq \Gamma (N\rightarrow \bar{l}\bar{\phi})\;,
\end{equation}
so that leptons and anti-leptons are produced in different amounts and a net 
$B-L$ asymmetry is generated\footnote{%
For this mechanism to work, the mass of the lightest Majorna neutrino should
be smaller than the postinflationary reheat temperature $T_{R}$. This is
difficult to reconcile in the context of the supergravity scenario, due to
an upper limit $T_{R}<10^{9}-10^{10}$ GeV arising from the thermal
production rate of gravitinos~\cite{R-therm}.}.

\strut

\textbf{3. The Lepton Leaking Mechanism}

\strut

We propose~\cite{bento01} an alternative mechanism of leptogenesis that is
based on scattering processes and lepton annihilation rather than decay
processes and lepton production. The idea is that lepton number and $CP$
violating interactions may exist with some hidden sector of new particles
which are not in thermal equilibrium with the ordinary world. The last
condition is automatically fulfilled if the two worlds only communicate via
gravity but other messengers may exist namely, superheavy gauge interaction
neutral particles, $N$, that mediate weak effective interactions between the
ordinary and hidden sectors at energies below their masses $M_{N}$. In this
scenario, 1) $L$ and $B-L$ are violated by one unit in reactions 
\begin{equation}
l\,\phi \,,\,\bar{l}\,\bar{\phi}\rightarrow X^{\prime }\;,
\end{equation}
that produce hidden particles, $X^{\prime }$, out of standard leptons and
Higgs, at temperatures much lower than the masses of the virtual mediators $%
N $. 2) These reactions are out-of-equilibrium i.e., are slow enough to not
bring the hidden sector and the reverse reactions into thermal equilibrium
which would destroy any lepton asymmetry. This implies that the initial
densities of the ordinary and hidden systems are different. In other words,
we assume that after inflation ends up, the reheat temperature of the
ordinary and hidden sectors are different, which can be achieved in certain
models \cite{KST,BDM}, and the latter is cooler or ultimately, completely
``empty''. The hidden sector starts to be ``slowly'' occupied by leaking of
the entropy from the ordinary to the hidden sector through $L$ violating
reactions. 3) $CP$ is violated in the effective interactions between leptons
and the hidden sector which originates from the $CP$ violation in
fundamental couplings between the messengers and both sectors. The result is
a $CP$ asymmetry in the average cross sections, 
\begin{equation}
\sigma (l\phi \rightarrow X^{\prime })\neq \sigma (\bar{l}\bar{\phi}%
\rightarrow X^{\prime })\;,
\end{equation}
and respective reaction rates, which means that leptons leak to the hidden
sector more (or less) effectively than antipletons so producing a net $B-L$
asymmetry. This is the reason we name it the leaking mechanism.

\strut

\textbf{4. The Mirror World Case}

\strut

The simplest model of this type can be described as follows. Consider the
lepton sector of the standard $SU(3)\times SU(2)\times U(1)$ model,
containing three generations of lepton doublets, $l_{i}=(\nu ,e)_{i}^{T}$, $%
i=1,2,3$, the standard Higgs doublet $\phi $, and some amount of heavy
singlet neutrinos $N_{a}$\footnote{%
Here and in the following we take all fermion states $\psi =l,N$ in the
left-handed basis and denote the (right-handed) antifermion states as $\bar{
\psi}=C\overline{\psi }^{T}$, where $C$ is the charge conjugation matrix.}.
Imagine now, that apart from the standard particles and interactions there
is an hidden sector with some gauge symmetry $G^{\prime }$ and fermion and
scalar fields $l_{k}^{\prime }$ and $\phi ^{\prime }$ that are singlets
under the standard model gauge symmetry group. The ordinary particles,
instead, carry zero quantum numbers under $G^{\prime }$. An interesting
candidate is a mirror sector, exact duplicate of the observable sector with
the same gauge symmetry and particle content, $G^{\prime }=SU(3)^{\prime
}\times SU(2)^{\prime }\times U(1)^{\prime }$. In any case, the heavy
singlet neutrinos $N_{a}$ can always play the role of messengers between
ordinary and hidden particles.

The Yukawa couplings have the form (charge conjugation matrix is omitted): 
\begin{equation}
h_{ia}l_{i}N_{a}\phi +h_{ka}^{\prime }l_{k}^{\prime }N_{a}\phi ^{\prime }+%
\frac{1}{2}M_{ab}N_{a}N_{b}+\mathrm{H.C.}\;.  \label{Yuk}
\end{equation}
The fields $l_{k}^{\prime }$ and $\phi ^{\prime }$ are doublets of $%
SU(2)^{\prime }$ and have zero lepton number contrary to the lepton fields $%
l_{i}$ ($L=1$) and singlet neutrinos $N_{a}$ (left-handed, $L=-1$) or $\bar{N%
}_{a}$ (right-handed, $L=+1$). The lepton number is violated by the $N_{a}$
Majorana masses ($\Delta L=2$) as well as by the $l_{k}^{\prime }N_{a}$
Yukawa couplings ($\Delta L=1$). Hence, we have a seesaw like scenario where
the heavy Majorana masses induce the masses of the ordinary active neutrinos
as much as the masses of the mirror neutrinos contained in $l_{k}^{\prime }$%
, sterile from our point of view, as well as the mixing terms between both
sectors~\cite{FV,BM}. This is the simplest sterile neutrino model that can
naturally explain why they may be light (with masses of the order of the
active neutrino masses) and have significant mixing with the ordinary
neutrinos.

Without loss of generality, the heavy neutrino mass matrix can be taken real
and parametrized as $M_{ab}=g_{ab}M$, $M$ being the overall mass scale and $%
g_{ab}$ order one constants. After integrating out the heavy neutrinos the
effective operators emerge as 
\begin{equation}
\frac{A_{ij}}{2M}l_{i}l_{j}\phi \phi +\frac{D_{ik}}{M}l_{i}l_{k}^{\prime
}\phi \phi ^{\prime }+\frac{A_{kn}^{\prime }}{2M}l_{k}^{\prime
}l_{n}^{\prime }\phi ^{\prime }\phi ^{\prime }+\mathrm{H.C.}\;,
\label{op-lpr}
\end{equation}
with coupling constant matrices given by $A=hg^{-1}h^{T}$, $A^{\prime
}=h^{\prime }g^{-1}h^{\prime T}$ and $D=hg^{-1}h^{\prime T}$. These
constants are complex as much as $h$ and $h^{\prime }$, to ensure the
existence of $CP$ violation.

\strut

\textbf{5. }$CP$\textbf{\ asymmetries}

\strut

In leading order, the total rate of lepton depeletion per unit of time and
existent lepton (with $L=1$) is given by $\Gamma _{1}=\sigma _{1}n_{b}$,
where $n_{b}\simeq (1.20/\pi ^{2})T^{3}$ is the thermal boson number density
per degree of freedom and $\sigma _{1}$ the total cross section of the $%
\Delta L=-1$ processes $l\phi \rightarrow \bar{l}^{\prime }\bar{\phi}%
^{\prime }$, 
\begin{equation}
\sigma _{1}=\sum \sigma (l\phi \rightarrow \bar{l}^{\prime }\bar{\phi}%
^{\prime })=\frac{Q_{1}}{8\pi M^{2}}\;,\qquad Q_{1}=\mathrm{Tr}%
(D\,D^{\dagger })\;,  \label{sigma1}
\end{equation}
where the sum is taken over all flavour and isospin initial and final states.

\begin{figure}[t]
\par
\begin{center}
\includegraphics*[width=80mm]{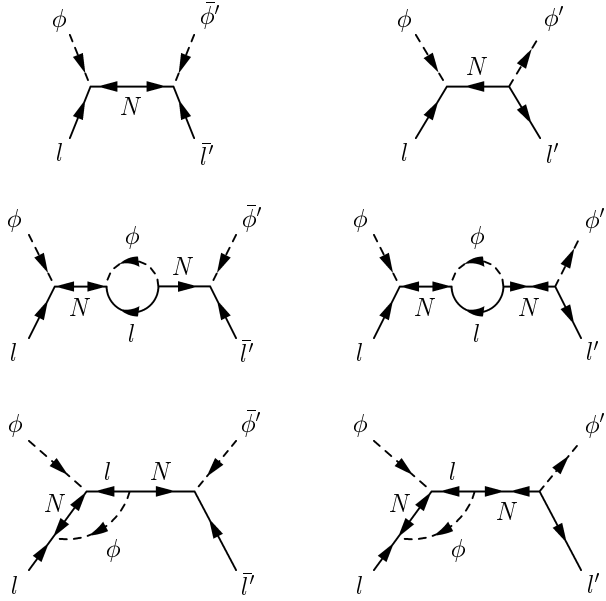}
\end{center}
\caption{Tree-level and one-loop diagrams contributing to the $CP$%
-asymmetries of $l\phi \rightarrow \bar{l}^{\prime }\bar{\phi}^{\prime }$
(left column) and $l\phi \rightarrow l^{\prime }\phi ^{\prime }$ (right
column). }
\label{fig1}
\end{figure}

\begin{figure}[t]
\begin{center}
\includegraphics*[width=80mm]{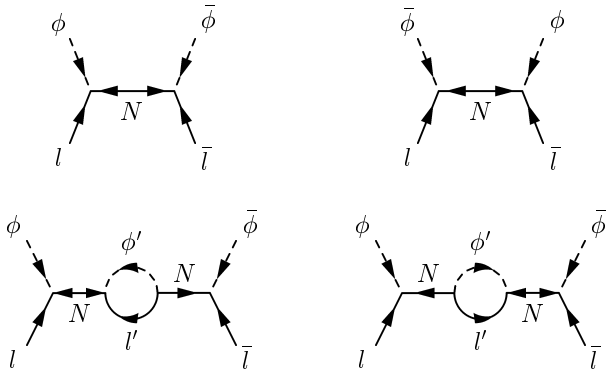} 
\end{center}
\caption{Tree-level and one-loop diagrams contributing to the CP-asymmetry
of $l\phi \rightarrow \bar{l}\bar{\phi}$.}
\label{fig2}
\end{figure}

The $CP$ asymmetries emerge from the interference between tree-level and
one-loop diagrams in much the same way as in the usual decay mechanisms. In
the case of $l\phi \rightarrow \bar{l}^{\prime }\bar{\phi}^{\prime }$ the
tree-level and one-loop diagrams are shown in the left column of Fig.\ (\ref
{fig1}). The tree-level amplitude goes as $M^{-1}$ and the radiative
correction as $M^{-3}$ hence, the $CP$ asymmetry of $l\phi \rightarrow \bar{l%
}^{\prime }\bar{\phi}^{\prime }$ versus $\bar{l}\bar{\phi}\rightarrow
l^{\prime }\phi ^{\prime }$ only appears at $M^{-4}$ order as shown in the
next equation ($\sqrt{s}$ is the c.m.\ energy). It turns out that the $l\phi
\rightarrow l^{\prime }\phi ^{\prime }$ reactions also present a $CP$
asymmetry at the same level of magnitude, actually exactly the same, despite that
their tree-level cross sections only contribute at $M^{-4}$ order. The
diagrams relevant for $l\phi \rightarrow l^{\prime }\phi ^{\prime }$ are
shown in the right column of Fig.\ (\ref{fig1}). Finally, one has to
consider the $\Delta L=2$ processes and their contribution to $B-L$
generation. We obtained the following $CP$ asymmetries: 
\begin{eqnarray}
&&\sigma (l\phi \rightarrow \bar{l}^{\prime }\bar{\phi}^{\prime })-\sigma (%
\bar{l}\bar{\phi}\rightarrow l^{\prime }\phi ^{\prime })=-\frac{1}{2}\Delta
\sigma _{CP}\;,  \nonumber \\
&&\sigma (l\phi \rightarrow l^{\prime }\phi ^{\prime })-\sigma (\bar{l}\bar{%
\phi}\rightarrow \bar{l}^{\prime }\bar{\phi}^{\prime })=-\frac{1}{2}\Delta
\sigma _{CP}\;,  \label{dsigma} \\
&&\sigma (l\phi \rightarrow \bar{l}\bar{\phi})-\sigma (\bar{l}\bar{\phi}%
\rightarrow l\phi )=\Delta \sigma _{CP}\;,  \nonumber
\end{eqnarray}
\begin{equation}
\Delta \sigma _{CP}=\frac{3s}{32\pi ^{2}M^{4}}J\;,  \label{dsigmacp}
\end{equation}
\begin{equation}
J=\mathrm{ImTr}[(h^{\prime \dagger }h^{\prime })g^{-2}(h^{\dagger
}h)g^{-1}(h^{\dagger }h)^{T}g^{-1}]\;.  \label{J}
\end{equation}
As proven below, the $\Delta L=2$ reactions $l\phi \longleftrightarrow \bar{l%
}\bar{\phi}$ and the $CP$ asymmetry associated with them are closely
related with the $\Delta L=1$ reactions by $CPT$ invariance. The diagrams
responsible for $CP$-violation in $l\phi \rightarrow \bar{l}\bar{\phi}$ and $%
\bar{l}\bar{\phi}\rightarrow l\phi $ are shown in Fig.\ \ref{fig2}.

\strut

\textbf{6. }$CPT$\textbf{\ invariance}

\strut

$CPT$ invariance implies that the total cross section for the scattering of
two particles is equal to the total cross section for the scattering of
their anti-particles if one takes an average (sum) over the initial spin
states. In particular, $\sigma (l\phi \rightarrow X)=\sigma (\bar{l}\bar{\phi%
}\rightarrow X)$, and the final relevant states are $\bar{l}^{\prime }\bar{%
\phi}^{\prime }$, $l^{\prime }\phi ^{\prime }$ and $\bar{l}\bar{\phi}$, $%
l\phi $. Taking into account that $CPT$ invariance also enforces $\sigma
(l\phi \rightarrow l\phi )=\sigma (\bar{l}\bar{\phi}\rightarrow \bar{l}\bar{%
\phi})$, one derives the following relation between the $CP$ asymmetries of $%
\Delta L=1$ and $\Delta L=2$ processes: 
\begin{equation}
\left[ \sigma (l\phi \rightarrow X^{\prime })-\sigma (\bar{l}\bar{\phi}%
\rightarrow X^{\prime })\right] +\left[ \sigma (l\phi \rightarrow \bar{l}%
\bar{\phi})-\sigma (\bar{l}\bar{\phi}\rightarrow l\phi )\right] =0
\label{CPT}
\end{equation}
where $X^{\prime }$ mean the states $\bar{l}^{\prime }\bar{\phi}^{\prime }$
and $l^{\prime }\phi ^{\prime }$. Then, the $CP$ asymmetries of both $\Delta
L=1$ and $\Delta L=2$ processes should cancel each other in agreement with
the direct calculation results shown in Eqs.\ (\ref{dsigma}). This
cancellation does not lead to a null lepton number variation because the $%
\Delta L$ variations are not identical to each other. In terms of the
reaction rate asymmetries ($\Gamma =\sigma \,n_{b}$) 
\begin{eqnarray}
\Gamma (l\phi &\rightarrow &X^{\prime })-\Gamma (\bar{l}\bar{\phi}%
\rightarrow X^{\prime })=\Delta \Gamma \;,  \nonumber \\
\Gamma (l\phi &\rightarrow &\bar{l}\bar{\phi})-\Gamma (\bar{l}\bar{\phi}%
\rightarrow l\phi )=-\Delta \Gamma \;,  \label{DeltaGama}
\end{eqnarray}
the rate of lepton number variation per unit of time and lepton particle
directly induced by these reactions is 
\begin{equation}
(-1)\Delta \Gamma +(-2)(-\Delta \Gamma )=+\Delta \Gamma \;.  \label{dldt}
\end{equation}
However, the lepton number is also violated by the electroweak instanton
processes in contrast with $B-L$, which is only violated in the above $%
\Delta L=1,2$ reactions. For that reason one can immediately establish the
net variation of $B-L$ as 
\begin{equation}
\frac{d}{dt}(B-L)=-\Delta \Gamma =\Delta \sigma _{CP}\,n_{b}\;,  \label{db-l}
\end{equation}
whereas the lepton number is determined from $B-L$ by the set of (standard
model) interactions that are in chemical equilibrium at each given
temperature and give the ratio between $L$ and $B-L$.

\strut

\textbf{7. }$B-L$\textbf{\ asymmetry}

\strut

One can evaluate the produced amount of lepton number of the universe in the
following way. Imagine that after inflation the inflaton field starts to
oscillate near its minimum and decays into ordinary and hidden particles
with different rates, thus giving rise to different reheat temperatures
for both sectors: $T_{R}^{\prime }<T_{R}$. For the very simplicity, one can
start to assume that the hidden sector is almost empty. As soon as the two
particle systems are produced, the interactions between them take over. The
only relevant reactions are the ones with ordinary particles in the initial
state and we assume that both $\Delta L=1$ and $\Delta L=2$ processes are
out of equilibrium. Hence, the $B-L$ asymmetry evolution is determined by
the $CP$ asymmetries shown in Eqs.\ (\ref{dsigma}) as follows: 
\begin{equation}
\frac{d\,n_{B-L}}{dt}+3H\,n_{B-L}=\Delta \sigma _{CP}\,n_{b}\, n_{f}\;,
\label{L-eq}
\end{equation}
where $n_{b}\simeq 0.122\,T^{3}$, $n_{f}=3/4\,n_{b}$ are the equilibrium
boson and fermion densities per degree of freedom. Noticing that the cross
section $CP$-asymmetry $\Delta \sigma _{CP}$ 
is proportional to the thermal average square c.m.\ energy $s\simeq
17\,T^{2} $ and $H=1/2\,t\propto T^{2}$, one integrates the above equation
from the reheat temperature $T_{R}$ to the low temperature limit obtaining
the produced $B-L$ asymmetry per unit of entropy 
\begin{equation}
Y_{B-L}=\left[ \frac{\Delta \sigma _{CP}\,n_{b}}{3\,H}\,Y_{f}\right]
_{R}\approx 2\times 10^{-3}\,J\frac{M_{Pl}T_{R}^{3}}{g_{*}^{3/2}\,M^{4}}\;,
\label{B-L}
\end{equation}
where $Y_{f}$ denotes the fermion number per unit of entropy and degree of
freedom ($H=1.66\,g_{*}^{1/2}T^{2}/M_{Pl}$ is the ordinary Hubble rate, $%
g_{*}$ the effective number of ordinary degrees of freedom, around $107$,
and $M_{Pl}\simeq 1.22\times 10^{19}$\textrm{\ GeV}).

In fact, the lepton number production starts before the reheat temperature
is achieved, as soon as the inflaton begins to decay and the particle
thermal bath is established. The calculation shows that the $B-L$ asymmetry
produced at temperatures above $T_{R}$ is $3/2$ the estimation (\ref{B-L}).

\strut

\textbf{8. Phenomenological constraints}

\strut

First of all one aims to match the observed baryon number asymmetry which
translates into the condition $Y_{B-L}\sim (2-3)\times 10^{-10}$. Another
immediate constraint is that the ordinary and hidden particles do not come
into thermal equilibrium with each other. The reason is twofold. First, it
would violate the limits on the number of extra light particle species at
the time of Big Bang Nucleosynthesis (BBN). Second, if the $\Delta L=1$
processes, mainly $l\phi \rightarrow \bar{l}^{\prime }\bar{\phi}^{\prime }$
and their charge conjugates were in equilibrium, they would erase $L$ and $%
B-L$. The other processes that may erase the lepton number are the $\Delta
L=2$ reactions $l\phi \leftrightarrow \bar{l}\,\bar{\phi}$, $%
l\,l\leftrightarrow \bar{\phi}\,\bar{\phi}$, $\phi \,\phi \leftrightarrow 
\bar{l}\,\bar{l}$. The out-of-equilibrium condition can be expressed as $%
K=\Gamma /H<1$, where $\Gamma =\gamma /n_{f}=\sigma \,n_{b}$ is the reaction
rate and $K$ represents the number of reactions per Hubble time. One obtains
for the $\Delta L=2$ and $\Delta L=1$ reactions respectively, 
\begin{eqnarray}
\gamma _{2} &=&\Gamma _{2}\,n_{f}\simeq \frac{3\,Q_{2}}{4\pi M^{2}}%
\,n_{b}n_{f}\;,\qquad \qquad Q_{2}=\mathrm{Tr}(A\,A^{\dagger })
\;,  \label{Gama2} \\
\gamma _{1} &=&\Gamma _{1}\,n_{f}=\frac{Q_{1}}{8\pi M^{2}}%
\,n_{b}n_{f}\;,\qquad \qquad Q_{1}=\mathrm{Tr}(D\,D^{\dagger })\;,
\label{Gama1}
\end{eqnarray}
the latter being dictated by Eq.\ (\ref{sigma1}).

$K_{1}=\Gamma _{1}/H$ and $K_{2}=\Gamma _{2}/H$ reach their maximum values
at the reheat temperature. It turns out that the condition $K_{1R}=\Gamma
_{1}/H\,(T_{R})<1$ is strong enough, first because after the Universe cools
down to the BBN epoch, the abundance of hidden particles (energy density $%
\rho ^{\prime }\approx 8K_{1R}\,\rho /g_{*R}$) translates into a number of
extra light neutrinos around $\Delta N_{\nu }\approx K_{1R}/2$, well inside
the present observational sensitivity, second because the mirror leptons
produced right after the reheating period are diluted into other flavours of
the hidden sector say, mirror quarks and bosons, as a result of gauge
interactions inside the hidden sector. Thus, the mirror leptons that are
actually available to produce a back reaction (rate $\gamma _{1}^{\prime }$)
and wash out the lepton number are only a fraction of the mirror leptons
produced (from ordinary particles) in $\Delta L=1$ reactions.
The back reaction rate relates to $\gamma _{1}$ as 
\begin{equation}
\frac{\gamma _{1}^{\prime }}{\gamma _{1}}=\frac{n_{b^{\prime }}n_{f^{\prime
}}}{n_{b}n_{f}}=\left( \frac{13 K_{1R}}{4g_{*R}^{\prime }}\right) ^{3/2}\; ,
\end{equation}
clearly suppressed by $g_{*R}^{\prime }$, the number of degrees of freedom of the hidden sector.

The out-of-equilibrium condition for the $\Delta L=2$ reactions, $%
K_{2}=\Gamma _{2}/H<1$, puts also an upper limit on the temperature, $%
T<T_{2} $. It is not absolutely necessary that the reheat temperature be
smaller than the decoupling temperature $T_{2}$. If $T_{2}<T_{R}$ then, $%
T_{2}$ marks the moment when the lepton number starts to be generated and
the final $B-L$ asymmetry is obtained by replacing $T_{2}$ for $T_{R}$ in
the expression (\ref{B-L}). Usually the rate $\Gamma _{2}$ is directly
related to the light neutrino masses which constitute the other
observational implication of any lepton number violating model. In our
enlarged framework two main possibilities can be considered. After standard
electroweak symmetry breaking the field $\phi ^{0}$ adquires an expectation
value $v\leq 174\mathrm{\ GeV}$. If on the contrary, the field $\phi
^{\prime }$ remains unbroken, the mirror leptons remain massless and do not
mix with the active neutrinos, whose mass matrix is given by 
\begin{equation}
m=-A\frac{v^{2}}{M}\qquad \Rightarrow \qquad \Gamma _{2}=\frac{\sum m_{\nu
}^{2}}{12\pi }\frac{T^{3}}{v^{4}}\;.  \label{m}
\end{equation}
Clearly, the rate $\Gamma _{2}$ and decoupling temperature $T_{2}$ are
determined by the active neutrino masses. They also set an upper bound on
the heavy neutrinos mass scale $M$. On the other hand, we assume that the
heavy neutrino masses are bigger than the reheat temperature $T_{R}$. This
is in strong contrast with the leptogenesis decay scenario. In particular,
we can account for the upper limit $T_{R}<10^{9}-10^{10}$ GeV which emerges
in the context of the supergravity models from the production rate of
gravitions \cite{R-therm}. Once we fix $T_{R}$ (or $T_{2}$ if $T_{R}>T_{2}$)
and the mass scale $M$ the required light neutrino mass spectrum and $B-L$
asymmetry indicate the range where the Yukawa couplings $h$ and $h^{\prime }$
entering in the matrices $A$ and $D$ must be.

The other possibilty is that both $\phi $ and $\phi ^{\prime }$ fields
adquire expectation values, $v$ and $v^{\prime }$. Then, the effective
operators of Eq.\ (\ref{op-lpr}) produce the mass matrix~\cite{BM} 
\begin{equation}
M_{\nu }=\left( 
\begin{array}{cc}
m & \mu \\ 
\mu & m^{\prime }
\end{array}
\right) =\frac{-1}{M}\left( 
\begin{array}{cc}
A\,v^{2} & D\,vv^{\prime } \\ 
D^{T}vv^{\prime } & A\,v^{\prime 2}
\end{array}
\right) \;.  \label{Mniu}
\end{equation}
The elements $\mu _{ik}$ mix the active and sterile (mirror) neutrino
sectors. It is important that the sterile neutrinos are not brought into
equilibrium through neutrino oscillations at low temperatures which would be
in conflit with the BBN limits on the number of extra light neutrinos. The
limit~\cite{dolgov81} on the sterile-active mixing angle is $\delta
m^{2}\sin ^{4}2\theta \lesssim 3\times 10^{-6}$\textrm{\ eV}$^{2}$. In view
of the mass scale involved in the atmospheric neutrino anomaly, $\delta
m^{2}\sim 3\times 10^{-3}$\textrm{\ eV}$^{2}$, that may be solved by
postulating an hierarchy between the $m$, $\mu $ and $m^{\prime }$ natural
scales namely, $h^{\prime } v^{\prime }/hv\sim \mu /m\sim m^{\prime }/\mu
\lesssim 10^{-2}$ which puts the upper limit $v^{\prime }\lesssim
17h/h^{\prime }\mathrm{\ GeV}$.

\strut

\textbf{9. Conclusions}

\strut

We propose a baryo- and leptogenesis mechanism in which the $B-L$ \
asymmetry is produced in the conversion of ordinary leptons into particles
of some depleted hidden sector. We studied the lepton number violating
reactions $l\phi \rightarrow l^{\prime }\phi ^{\prime },\,\bar{l}^{\prime }%
\bar{\phi}^{\prime }$ mediated by the usual heavy Majorana neutrinos $N$ of
the seesaw mechanism~\cite{seesaw}, $l$ and $\phi $ being the ordinary
lepton and Higgs doublets and $l^{\prime }$, $\phi ^{\prime }$ their
``sterile'' counterparts. This mechanism can operate even if the reheat
temperature is smaller than the $N$ Majorana masses, in which case the usual
leptogenesis mechanism through the $N$ decays is ineffective. The reheat
tempearture can be as low as $10^{9}$ GeV or less. In particular, we can
account for the upper limit $T_{R}<10^{9}-10^{10}$ GeV which emerges in the
context of the supergravity models from the production rate of gravitions~%
\cite{R-therm}. This is in strong contrast with the leptogenesis decay
scenario~\cite{fukujita86,luty92,buchmuller99}.

The mechanism we propose can be realized in different ways as it does not
crucially depend on any particular model. The main idea is that there exists
some hidden sector of new particles which interact very weakly with the
ordinary particles. Such interactions must be weak enough to not put the
hidden sector in thermal equilibrium with the ordinary sector in the Early
Universe but, on the other hand, must and can be strong enough to cause a
leakage of ordinary leptons (baryons) in $L$ ($B$) violating collision
reactions capable of producing the desired very small $B-L$ ($L$ and $B$)
asymmetry that is needed. This is achieved with $CP$ violating couplings
causing an asymmetric leakage of fermions and antifermions.

In the example we worked out in detail, the communicators are $\Delta L=1$
effective interactions mediated by the heavy right-handed neutrinos $N$ of
the seesaw mechanism. But in principle other kind of particles can play such
a messenger role. The hidden sector contains typically fermion and scalar
fields with their own gauge interactions under some gauge symmety group $%
G^{\prime }$, but are singlets under the standard model gauge group $%
G=SU(3)\times SU(2)\times U(1)$, while the ordinary particles are instead
singlets under $G^{\prime }$. The messenger particles are singlets under
both gauge groups $G$ and $G^{\prime }$.

The interesting hidden sector candidate can be a mirror sector with the same
gauge symmetry $G^{\prime }$ and identical particle content. The old idea of
such a mirror sector that is the exact duplicate of our visible world \cite
{LY} has attracted a significant interest over the last years motivated in
particular by the problems of neutrino physics~\cite{FV,BM} and other
problems in particle physivs and cosmology~\cite{BDM,LY,BCV}. The basic
concept is to have a theory given by the product $G\times G^{\prime }$ of
two identical gauge factors with identical particle contents, which could
naturally emerge e.g.\ in the context of $E_{8}\times E_{8}^{\prime }$
superstring theories. However, in the more general case $G^{\prime }$ could
be any gauge symmetry group.

An important point~\cite{bento01} is that  the same mechanism that 
produces the lepton number in the ordinary Universe, 
can also produce the lepton prime asymmetry in 
the hidden sector. 
Then, if it contains also some baryon like
heavier particles, it can provide a type of 
self-interacting dark matter.
In the context of our lepto-baryogenesis mechanism
 $n_{B-L} \sim n^\prime_{B-L}$ and so the 
mirror baryon number density should be comparable to the ordinary 
baryon density, $\Omega_B^\prime \sim \Omega_B$. 
Another important fact is,
the magnitude of the produced $B-L$, Eq.\ (\ref{B-L}), 
strongly depends on the reheating temperature hence, 
a larger $B-L$ is produced in 
the places where the temperature is higher. 
In the cosmological context, this would lead to 
a situation where apart from the adiabatic density/temperature 
perturbations, there also emerge correlated isocurvature 
fluctuations with variable $B$ and $L$ 
which could be tested with the future 
data on the CMB anisotropies.

\strut

We thank FCT for the grant CERN/P/FIS/40129/2000. The work of Z. B. was
partially supported by the MURST research grant ``Astroparticle Physics''.

\newpage


\begin{thebibliography}{99}
\bibitem{sakharov67}  A. D. Sakharov, JETP Lett. \textbf{5}, 24 (1967).

\bibitem{kuzmin85}  V. A. Kuzmin, V. A. Rubakov and M. E. Shaposhnikov,
Phys. Lett. {\textbf{155B}}, 36 (1985).

\bibitem{khlebnikov88}  S. Yu. Khlebnikov and M. E. Shaposhnikov, Nucl.
Phys. B \textbf{308}, 885 (1988).

\bibitem{harvey90}  J. A. Harvey and M. S. Turner, 
Phys. Rev. D {\textbf{42}}, 3344 (1990).

\bibitem{Dolgov}  For a review, see 
A.\ D.\ Dolgov, Phys. Rep. \textbf{222}, 309 (1992); 
A. Riotto and M. Trodden, Annu. Rev. Nucl. Part. Sci. \textbf{49}, 
35 (1999);  
see also ref.~\cite{buchmuller99}.

\bibitem{fukujita86}  M. Fukujita and T. Yanagida, 
Phys. Lett. B {\textbf{174}}, 45 (1986); Phys. Rev. D \textbf{42}, 1285 (1990).

\bibitem{luty92}  M. A. Luty, Phys. Rev. D \textbf{45}, 455 (1992);
M. Pl\"{u}macher, Z. Phys. C \textbf{74}, 549 (1997);
M. Flanz and E. A. Paschos, Phys. Rev. D \textbf{58}, 113009 (1998).

\bibitem{buchmuller99}  W. Buchm\"{u}ller and M. Pl\"{u}macher, Phys. Rep. 
\textbf{320}, 329 (1999) and Int. J. Mod. Phys. A {\textbf{15}}, 5047 (2000);
A. Pilaftsis, Int. J. Mod. Phys. A \textbf{14}, 1811 (1999).

\bibitem{seesaw}  M. Gell-Mann, P. Ramond and R. Slansky, 
in \emph{Supergravity}, Proceedings of the Workshop, Stony Brook, New York, 1979, edited by P. van Nieuwenhuizen and D. Freedman (North-Holland, Amsterdam,
1979); 
T. Yanagida, in \emph{Proceedings of the Workshop on Unified Theories and Baryon Number in the Universe}, Tsukuda, Japan, 1979, edited by A. Sawada and A. Sugamoto
(KEK Report No. 79-18, Tsukuda, 1979); 
R. N. Mohapatra and G. Senjanovi\'{c}, Phys. Rev. Lett. \textbf{44}, 912 (1980).

\bibitem{R-therm} J.\ R. Ellis, A.\ D. Linde and D.\ V. Nanopoulos, 
Phys. Lett. \textbf{118B}, 59 (1982);
D.\ V. Nanopoulos, K.\ A. Olive and M. Srednicki,  
\emph{ibid.} \textbf{127B}, 30 (1983).

\bibitem{bento01}  L. Bento and Z. Berezhiani, 
Phys. Rev. Lett. \textbf{87}, 231304 (2001).

\bibitem{KST}  E. Kolb, D. Seckel, M. Turner, Nature 514, \textbf{415}
(1985).

\bibitem{BDM} Z.\ G. Berezhiani, A.\ D. Dolgov and R.\ N. Mohapatra,   
Phys. Lett. B \textbf{375}, 26 (1996);
Z.\ G. Berezhiani, Acta Phys. Pol. B \textbf{27}, 1503 (1996);
V.\ S. Berezinsky, A. Vilenkin, Phys. Rev. D \textbf{62}, 083512 (2000).

\bibitem{FV}  R. Foot and R. R. Volkas, Phys. Rev. D \textbf{52}, 6595
(1995); For earlier work, R. Foot, H. Lew and R. R. Volkas, Mod. Phys. Lett.
A \textbf{7}, 2567 (1992).

\bibitem{BM}  Z. G. Berezhiani and R. N. Mohapatra, 
Phys. Rev. D \textbf{52}, 6607 (1995); 
For earlier work, E. Kh. Akhmedov, Z. G. Berezhiani and G.
Senjanovi\'{c}, Phys. Rev. Lett. \textbf{69}, 3013 (1992).

\bibitem{dolgov81}  A.\ D. Dolgov, Yad. Fiz. \textbf{33}, 1309 (1981)
[{Sov. J. Nucl. Phys.} {\textbf{33}}, 700 (1981)]; 
R. Barbieri and A. Dolgov, {Phys. Lett. B} {\textbf{237}}, 440 (1990); 
{Nucl. Phys. B} {\textbf{349}}, 743 (1991); 
K. Kainulainen, {Phys. Lett. B} {\textbf{244}}, 191 (1990); 
K. Enqvist, K. Kainulainen and M. Thomson, 
{Nucl. Phys. B} {\textbf{373}}, 498 (1992); 
J.\ M. Cline, {Phys. Rev. Lett.} {\textbf{68}}, 3137 (1992); 
X. Shi, D. N. Schramm and B. D. Fields, {Phys. Rev. D} {\textbf{48}}, 2563 (1993).

\bibitem{LY} T.\ D. Li and C.\ N. Yang, Phys. Rev. \textbf{104}, 254 (1956);
I.\ Yu. Kobzarev, L.\ B. Okun, I.\ Ya. Pomeranchuk, 
Yad. Fiz. \textbf{3}, 1154 (1966) 
[Sov. J. Nucl. Phys. \textbf{3}, 837 (1966)]; 
M. Pav\v{s}i\v{c}, Int. J. Theor. Phys. \textbf{9}, 229 (1974);
S.\ I. Blinnikov and M.\ Yu. Khlopov, Astron. Zh. \textbf{60}, 632 (1983)
[Sov. Astron. \textbf{27}, 371 (1983)];
M.\ Yu. Khlopov \emph{et al.}, Astron. Zh. \textbf{68}, 42 (1991)
[Sov. Astron. \textbf{35}, 21 (1991)];
R. Foot, H. Lew and R.\ R. Volkas, Phys. Lett. B \textbf{272}, 67 (1991);
Mod. Phys. Lett. A \textbf{7}, 2567 (1992);
H.\ M. Hodges, Phys. Rev. D \textbf{47}, 456 (1993);
Z.\ K. Silagadze, Phys. At. Nucl. \textbf{60}, 272 (1997); 
Z. Berezhiani, Phys. Lett. B \textbf{417}, 287 (1998).


\bibitem{BCV}  Z. Berezhiani, D. Comelli and F.\ L. Villante, 
Phys. Lett. B \textbf{503}, 362 (2001).


\end{thebibliography}
\end{document}